\documentclass[onecolumn,preprintnumbers,pre]{revtex4}
\usepackage{graphicx, subfigure}
\usepackage{amssymb, amsmath,amssymb,amsfonts}
\usepackage{amsthm,mathrsfs,amsopn}
\usepackage{dcolumn}
\usepackage{bm}
\usepackage{color}
\usepackage[utf8]{inputenc}

\theoremstyle{plain} 

\newtheorem{theorem}{Theorem}

\newtheorem{remark}[theorem]{Remark}

\begin{document}

\title{Random walks and community detection in hypergraphs}

\author{Timoteo Carletti$^1$}
\affiliation{$^1$naXys, Namur Institute for Complex Systems, University of Namur, Belgium}
\author{Duccio Fanelli$^2$}
\affiliation{$^2$Universit\`a degli Studi di Firenze, Dipartimento di Fisica e Astronomia, CSDC and INFN, via G. Sansone 1, 50019 Sesto Fiorentino, Italy}
\author{Renaud Lambiotte$^3$}
\affiliation{$^3$Mathematical Institute, University of Oxford, UK}

\begin{abstract}
We propose a {one parameter} family of random walk processes on hypergraphs, where a parameter biases the dynamics of the walker towards hyperedges of low or high cardinality. We show that {for each value of the parameter the resulting} process defines its own {hypergraph} projection on a weighted network. We then explore the differences between them by considering the community structure associated to each random walk process. To do so, we generalise the Markov stability framework to hypergraphs and test it on artificial and real-world hypergraphs.
\end{abstract}

\maketitle

\section{Introduction}
\label{sec:intro}

In the last two decades, networks have emerged as a powerful framework to model and study complex systems~\cite{AlbertBarabasi,BLMCH}, providing a rich set of tools and methods that can be used independently of the nature of their {constituting} components~\cite{Newmanbook,Barabasibook,Latorabook}.  At the core of network science, there is the interchange between structure and dynamics~\cite{Castellanoreview,arenasreview,BBV2008}, as the topology imposes constraints of how dynamical processes propagate between nodes and, reversely, dynamical processes are at the core of several algorithms to extract information from the underlying structure~\cite{schaub2019structured}. An important context where this interplay is essential is the modular, or community, structure of networks. On the one hand, several works have shown that the presence of communities slows down diffusive processes on networks, in particular random walk processes~\cite{Rednerbook}. Here, communities are understood as groups of densely connected nodes, leading to the presence of bottlenecks between communities. Important concepts include the Cheeger inequality, providing relations between the mixing time, e.g. time for a random walk process to relax to stationarity, and the notion of conductance, measuring the bottlenecks in networks. On the other hand, random walk processes are at the heart of several algorithms to extract communities in networks. Community detection~\cite{Fortunato2012}  is a central element in the toolkit of network science, allowing to produce coarse-grained representation of large-scale networks~\cite{SDLB,SOCBDLB} {and thus to identify groups of nodes whose behaviour is relevant for the process under study}. Methods based on random walks include the Map Equation~\cite{rosvall2008} and Markov stability~\cite{DelvenneEtAl2010,LambiotteEtAl2014}, exploiting the fact that good communities tend to capture random walkers for long times before they can escape them. 

Networks make assumptions about the structure of interacting systems, often implicitly, and recent research has questioned the relevance of these assumptions in real-world data, leading to the concept of higher-order networks~\cite{BGL,LRS2019}. In networks, for instance, the building blocks for interactions are pairwise interactions, and the whole network is formed by combining such pairwise interactions. In contrast, many interacting systems are made of interactions that involve more {than} two nodes and can not be decomposed further{, i.e. they act as a whole}. A canonical example is collaboration networks, where groups of authors interact to produce research papers~\cite{PPV2017,CarlettiEtAl2020}{: the emerging results are a product of the group rather than reflecting pairwise interexchanges.} More generally, systems characterised by multibody interactions  abound in a variety of scientific domains, from functional brain networks~\cite{petri2014homological,LEFGHVD2016} to  protein interaction networks~\cite{estradaJTB} and ecology~\cite{GBMSA}. 

The inadequacy of standard networks to model multibody interactions motivates the  development of appropriate higher-order models, often enriching and generalising the standard network paradigm. The two most popular approaches are  simplicial complexes~\cite{DVVM,BC,PB} and  hypergraphs~\cite{berge1973graphs,estrada2005complex,GZCN}. 
Simplicial complexes are a model from topological data analysis, whose aim is to characterise the shape of data, in terms of the presence of holes between the points, and have been proposed as generalisations of networks with applications in epidemic spreading~\cite{BKS2016,IPBL} and synchronisation~\cite{LCB2020,GdPGLRCFLB2020}. 
The focus of this paper is on hypergraphs, a domain that has a long tradition in graph theory~\cite{berge1973graphs} but whose impact on dynamics has only been considered more recently. Recent works include applications to 
social contagion model~\cite{de2019social,ATM2020}, the modelling of random walks~\cite{CarlettiEtAl2020}, the study of synchronisation~\cite{Krawiecki2014,MKJ2020,CarlettiEtAl2020b}, diffusion~\cite{ATM2020}, non-linear consensus \cite{neuhauser2020multibody} and the emergence of Turing patterns~\cite{CarlettiEtAl2020b}. Hypergraphs constitute a powerful and flexible paradigm, as they encode interactions by hyperedges, defined as groups of arbitrary size between nodes.  In situations when all the hyperedges have size 2, hypergraphs reduce to standard networks.  Hypergraphs have several advantages, as they allow to efficiently handle very large hyperedges and, even more importantly, a heterogenous distribution of hyperedges' sizes. Moreover, the information on the high-order structure of the embedding support are stored in a matrix whose dimension depends only on the number of nodes~\cite{CarlettiEtAl2020,ATM2020}  thus avoiding the use of tensors.

The main purpose of this paper is to explore the interplay between dynamics and structure in hypergraphs. More specifically, we will generalise the notion of Markov stability. As we will discover, different types of random walk processes can be defined on hypergraphs, each leading to different quality functions and different partitions into communities. This flexibility originates from the fact that hyperedges are characterised by a {feature}, their size, and that a choice has to be made to model how this {feature} affects the diffusion. Classical studies in graph theory often assume that the hypergraphs are uniform, that is the hypergedges all have the same size~\cite{lu2011high,HL2019}.
The first random walk Laplacian defined for  general hypergraphs can  probably be traced back to the seminal paper  of Zhou and collaborators~\cite{zhou2007learning}, where each hyperedge is endowed with an arbitrary weight, acting as a veritable bias to the walkers dynamics. While this weight is often considered to be a free parameter that can be chosen a priori, it may naturally emerge in certain models, such as in~\cite{CarlettiEtAl2020} where the transition rates of the  process  are linearly biased by the size of the hyperedges, i.e. a walker follows a hyperedge proportionally to its size. 
In general, different choices for this weight, and the resulting biases on the random walk trajectories, lead to different transition matrices. Our main result is to investigate and to characterise how these differences translate into different communies in hypergraphs.

This paper is organised as follows. In section \ref{sec:rwh}, we define random walks on hypergraphs and introduce a one-parameter family of processes, extending  previous models. In section \ref{sec:mcomm}, we generalise the concept of Markov stability to hypergraphs and, in section \ref{sec:ex}, we explore the impact of the parameter of the random walk process on the uncovered communities, looking at both artificial and real-world networks. Finally, in section \ref {sec:conc}, we conclude and discuss the implications of our work.

\section{Hypergraphs and random walks}
\label{sec:rwh}
Let us consider an hypergraph $\mathcal H(V,E)$, where $V=\{v_1,\dots,v_n\}$ denotes the set of $n$ nodes and $E=\{E_1,\dots,E_m\}$ the set of $m$ hyperedges, that is for all $\alpha=1,\dots,m$: $E_\alpha\subset V$, i.e. an unordered collection of vertices. Note that if $E_\alpha=\{u,v\}$, i.e. $|E_\alpha|=2$, then the hyperedge is actually a ``standard'' edge denoting a binary interaction among $u$ and $v$. If all hyperedges have size $2$, the hypergraph is thus a network.
The hypergraph can be encoded by its {\em incidence matrix}  $e_{i \alpha}$, where we adopt the convention of using roman indexes for nodes and greek ones for hyperedges
\begin{equation}
\label{eq:incid}
e_{i \alpha}=\begin{cases} 1 &\text{$v_i\in E_{\alpha}$}\\
0 & \text{otherwise}\, .
\end{cases}
\end{equation}

A standard procedure to construct the $n\times n$ adjacency matrix of the hypergraph is $\mathbf{A}=\mathbf{e}\mathbf{e}^{T}$, whose entry $A_{ij}$ represents the number of hyperedges containing both nodes $i$ and $j$. Note that it is often customary to set to zero the diagonal elements of the adjacency matrix. Let us also define the $m\times m$ hyperedges matrix $\mathbf{B}=\mathbf{e}^{T}\mathbf{e}$, whose entry $B_{\alpha \beta}$ counts the number of nodes in $E_{\alpha}\cap E_{\beta}$. $\mathbf{B}$ can be seen as the (weighted) adjacency matrix of the dual hypergraph, i.e. where hyperedges of the original hypergraph become nodes of the new structure and two nodes are connected by a weighted link counting how many nodes (in the original hypergraph) are shared by the two hyperedges, namely $B_{\alpha\beta}$. Note that a similar construction has been proposed in~\cite{Evans_2010} to extract a $n$-clique graph from a network, namely a network whose nodes are the $n$-clique of the original one and whose nodes are connected if the $n$-clique share at least one node (in the original network). The main difference in the present case is that hyperedges can have an heterogeneous size distribution and thus provide a more flexible framework.

We can define a random walk process on a hypergraph as follows. The agents are located on the nodes and hop between nodes at discrete times. In a general setting, the walkers may give more or less importance to hyperedges depending on their size, {a choice} that may then bias  their moves~\cite{CarlettiEtAl2020}. This process is captured by the weighted adjacency matrix 
\begin{equation}
\label{eq:khij}
K^{(\tau)}_{ij}=\sum_{\alpha}(B_{\alpha \alpha}-1)^{\tau}e_{i\alpha}e_{j\alpha}\quad\forall i\neq j \quad\text{and $K^{(\tau)}_{ii}=0$}\, ,
\end{equation}
where $\tau$ is a real parameter whose role will be discussed in the following{; note that $B_{\alpha \alpha}-1$ denotes the number of nodes in the hyperedge $E_\alpha$ available to the walker, i.e. discarding the node where she is sitting initially}. It is worth emphasising that in principle we could have used  a generic monotone function $f$ of the hyperedge size, in defining the above quantities. For illustrative purposes we have here chosen to limit the analysis to the relevant, although particular, setting of a power law bias. The transition probabilities of the examined process are then obtained by normalising the columns of the weighted adjacency matrix
\begin{equation}
\label{eq:Tij4}
T^{(\tau)}_{ij}=\frac{K^{(\tau)}_{ij}}{\sum_{{\ell\neq i}} K^{(\tau)}_{i\ell}}\quad\forall i\neq j\quad\text{and $T^{(\tau)}_{ii}=0$}\, .
\end{equation}
Note that a ``lazy" process could have been defined by assuming that walkers are able to stay put on a  node in one step, for instance by taking $K^{lazy,(\tau)}_{ij}=\sum_{\alpha}(B_{\alpha \alpha})^\tau e_{i\alpha}e_{j\alpha}$.

This general definition covers several existing models of random walks on hypergraphs.
For $\tau=1$, we  get the random walk defined in~\cite{CarlettiEtAl2020} while for $\tau=-1$ we obtain the one introduced by Zhou~\cite{zhou2007learning}, as
\begin{equation*}
 T^{(-1)}_{ij}=\frac{K^{(-1)}_{ij}}{\sum_{{\ell\neq i}} K^{(-1)}_{i\ell}}=\frac{1}{\sum_{{\ell\neq i}} K^{(-1)}_{i\ell}} \sum_{\alpha}\frac{e_{i\alpha}e_{j\alpha}}{(B_{\alpha \alpha}-1)}\, ,
\end{equation*}
and
\begin{equation*}
\sum_{{\ell\neq i}} K^{(-1)}_{i\ell}=\sum_{{\ell\neq i}}\sum_{\alpha}\frac{e_{i\alpha}e_{\ell\alpha}}{(B_{\alpha \alpha}-1)}=\sum_{\alpha}{e_{i\alpha}}=k_i\, ,
\end{equation*}
where we used the fact $\sum_{{\ell\neq i}}e_{\ell\alpha}$ equals the number of nodes in $E_\alpha$ without node $i$, that is $(B_{\alpha \alpha}-1)$, which thus simplifies with the denominator. The last equality defines $k_i$ to be number of hyperedges incident to node $i$, a quantity that is simply the node degree in the case of networks. In conclusion,
\begin{equation}
\label{eq:Tmeno1}
 T^{(-1)}_{ij}=\sum_{\alpha}\frac{e_{i\alpha}e_{j\alpha}}{k_i(B_{\alpha \alpha}-1)}\, 
\end{equation}
 has the following interpretation: the walker sitting on node $i$ choses uniformly at random one hyperedge among the incident ones, i.e. with probability $1/k_i$, and then it selects a node belonging to the latter, again uniformly at random, i.e. with probability $1/(B_{\alpha \alpha}-1)$. 

\begin{figure}[ht]
\centering
\includegraphics[scale=0.2]{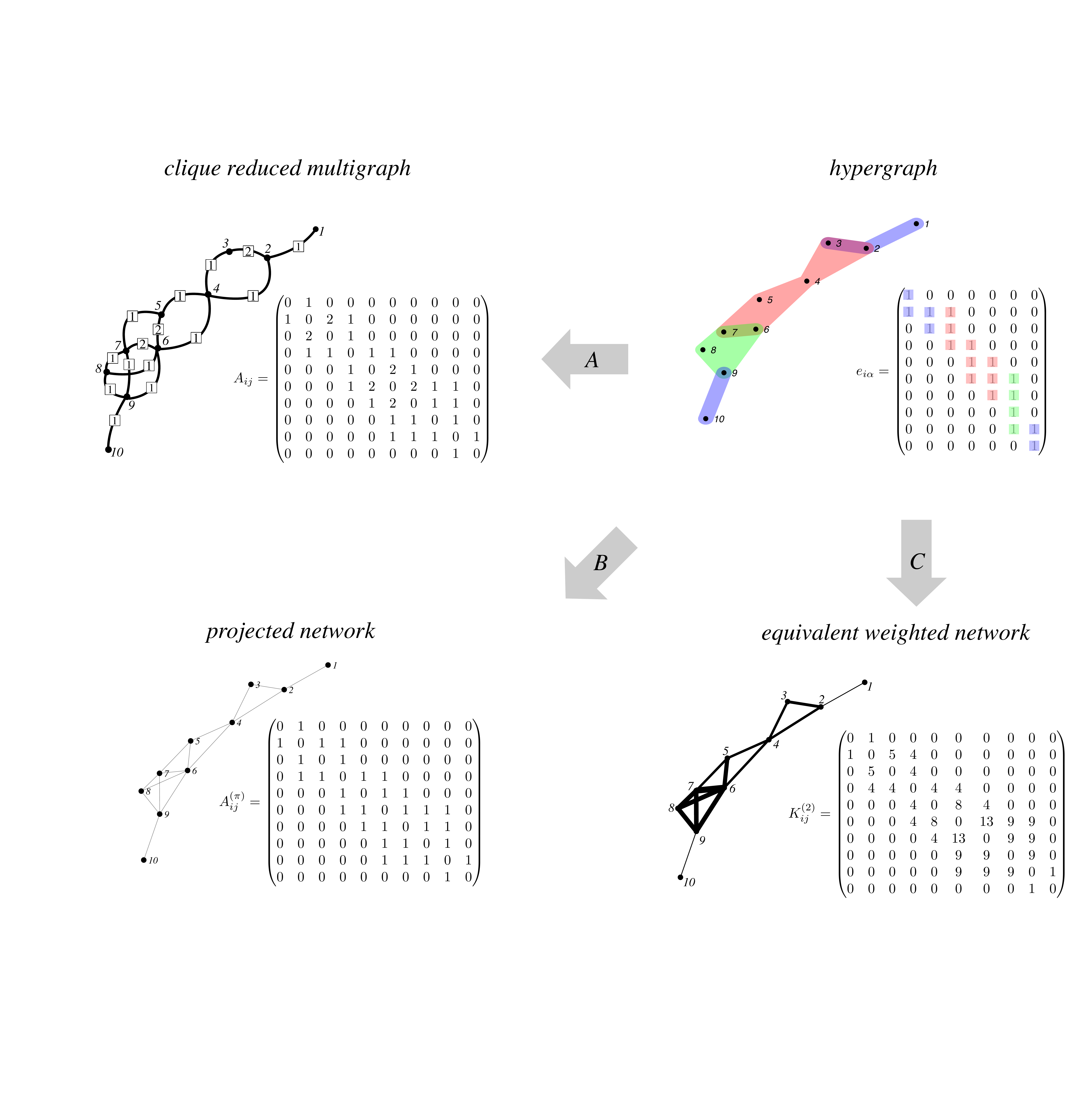}
\vspace{-2.5cm}
\caption{\textbf{Hypergraph and their projection}. The top right panel shows a hypergraph, where  hyperedges of different sizes are shown in different colours.  Its structure is entirely encoded in the incidence matrix $e_{i\alpha}$ where, for ease of visualisation, we use the same colour code. 
Different projections can be constructed from the same hypergraph, each leading to a different weighted graph. In the clique reduced multigraph (top left panel), links between two nodes carry a weight proportional to the number of hyperedges including these two nodes. Alternatively, 
 the so-called projected network is a binarised version of the clique reduction multigraph, so that a link exists between two nodes if they belong to at least one hyperedge. Finally, the family of projections $\mathbf{K}^{(\tau)}$ defined in this paper, illustrated for the case $\tau=2$ at the bottom right, gives more or less weight to edges  depending on the size of the hyperedges from which  they are built.}
\label{fig:hypergraph}
\end{figure}

The case $\tau=0$ returns a random walk on the so called {\em clique reduced multigraph}. The latter is a multigraph where each pair of nodes is connected by a number of edges equal to the number of hyperedges containing that pair in the hypergraph (see top left panel of Fig.~\ref{fig:hypergraph}). In that case, the transition matrix simplifies into
\begin{equation*}
 T^{(0)}_{ij}=\frac{K^{(0)}_{ij}}{\sum_{{\ell\neq i}} K^{(0)}_{i\ell}}=\frac{A_{ij}}{\sum_{{\ell\neq i}} A_{i\ell}} \, ,
\end{equation*}
where we used the definition of the hyperadjacency matrix $A_{ij}=\sum_{\alpha}e_{i\alpha}e_{j\alpha}$.
Let us observe that the clique reduced multigraph is different from the {\em projected network} obtained by associating to each hyperedge a clique of the same size;  the projected network can be interpreted as the unweighted (binarised) version of the {\em clique reduced} multigraph (see bottom left panel of Fig.~\ref{fig:hypergraph}).

From the definitions \eqref{eq:khij} and~\eqref{eq:Tij4}, it is clear that hyperedges with larger sizes will dominate and, thus, set the fate of the random process for large values of $\tau$. 
When $\tau$ is very negative, in contrast,  hyperedges with a  small size will drive the random walk process. As such, $\tau$ can be pictured as a {\em size bias parameter} which dictates the importance of hyperedges depending on their size.

Given the transition probability~\eqref{eq:Tij4}, a continuous-time random walk can be defined by
\begin{equation*}
\dot{p}_i(t)=\sum_j p_j(t)T^{(\tau)}_{ji} - \sum_j p_iT^{(\tau)}_{ij}\, ,
\end{equation*}
where ${p}_i(t)$ is the probability of finding the walker on node $i$ at time $t$. Note that $\mathbf{p}=(p_1,\dots,p_n)$ is a row vector, as often assumed once dealing with Markov processes. 
Using $\sum_j T^{(\tau)}_{ij}=1$, the latter can be rewritten
\begin{equation*}
\dot{p}_i=\sum_j p_j(T^{(\tau)}_{ji} - \delta_{ij})=-\sum_j p_jL^{(\tau)}_{ji}\, ,
\end{equation*}
where
\begin{equation}
\label{eq:rwLap}
L^{(\tau)}_{ij}=  \delta_{ij}-T^{(\tau)}_{ij} = \delta_{ij}-\frac{K^{(\tau)}_{ij}}{\sum_{\ell \neq i} K^{(\tau)}_{i\ell}}\, ,
\end{equation}
is a  {{\em random walk}} Laplacian generalising that of standard networks. Note that the standard Laplacian is recovered in the case $|E_\alpha|=2$ for all $\alpha$.

As this Laplacian, and its associated random walk process, can be interpreted as a standard random walk on the weighted {undirected network encoded by the} symmetric adjacency matrix $K^{(\tau)}_{ij}$, standard results naturally generalise and we find that the stationary state is
\begin{equation}
p_j^{(\infty)}=\frac{d_j^{(\tau)}}{\sum_{\ell} d_{\ell}^{(\tau)}}\,,
\label{eq:statnorm}
\end{equation}
where $d_j^{(\tau)}=\sum_{\ell\neq j}K_{j\ell}^{(\tau)}$ is the strength of node $j$ in the weighted graph. This latter quantity is an immediate generalisation of the standard node degree, in a direction which enables to account for the existence of different types of hyperedges in the system. Note that this interpretation of  \eqref{eq:rwLap} as a random walk  on the weighted projected network $K_{ij}^{(\tau)}$ only holds for that projection, and not for other projections such as the clique reduced multigraph (see Figure 1). 
This observation is critical, as it allows us to easily adapt standard results from network science  to the hypergraph framework passing by the weighted projected network $K_{ij}^{(\tau)}$.

\begin{remark}[Connection with bipartite works]
 As appears from the incidence matrix (\ref{eq:incid}), hypergraphs can be equivalently represented by bipartite newtorks, where one set of nodes would be the nodes of the hypergraph, and the other set would be the hyperedges.  Projecting hypergraphs thus finds connections with the well-studied problem of projecting bipartite networks. Important choices include 
 \begin{equation*}
 w_{ij}^{Newman}=\sum_\alpha \frac{e_{i\alpha}e_{j\alpha}}{B_{\alpha\alpha}-1}\, ,
\end{equation*}
proposed to build scientific collaboration networks \cite{Newman2001} and clearly equivalent to 
 $K_{ij}^{(-1)}$  in Eq.~\eqref{eq:khij}. Other choices include \cite{ZRMZ2007}
 \begin{equation*}
  w_{ij}^{Zhou}=\sum_{\alpha}\frac{e_{i\alpha}e_{j\alpha}}{k_iB_{\alpha \alpha}}\, ,
\end{equation*}
where the full size of the hyperedge is now taken into account, which corresponds to $\tau=-1$ with a lazy walker.
The formulas \eqref{eq:khij} and~\eqref{eq:Tij4} are thus a natural generalisation of projection methods for bipartite networks as well.
\end{remark}

\section{Markov stability and Community detection in hypergraphs}
\label{sec:mcomm}

Different algorithms for community detection are based on random walk processes and, more specifically, on the intuition that walkers should stay for long times inside good communities before escaping them. As we have shown in the previous section, the size bias parameter allows us to bias the trajectories of random walkers or, equivalently, to give more or less weight to certain edges in the projected graph. This observation motivates the use of random walkers with different values of $\tau$ in order to search for communities giving more, or less, importance to edges belonging to large hyperedges. This process should provide us with communities that would differ from those obtained in the standard  clique reduced multigraph, except for the particular case $\tau=0$. 

In the following, we will search communities by generalising the flexible framework of Markov stability~\cite{DelvenneEtAl2010,LambiotteEtAl2014}. Let us consider a partition of the nodes of a hypergraph into $c$ (non overlapping) communities, encoded by the $n \times c$ indicator matrix $\mathbf{C}$ with $C_{ij} \in \{0, 1\}$, where a $1$ denotes that node $i$ belongs to community $j$ and $0$ otherwise. Given a partition $\mathbf{C}$, one can define the Markov stability
\begin{equation}
\label{eq:r}
r(t;\mathbf{C}):=\min_{0\leq s\leq t}\mathit{tr} \left[ \mathbf{C}^T\left(\mathbf{\Pi} e^{-s\mathbf{L}^{(\tau)}} - (\mathbf{p}^{(\infty)})^T\mathbf{p}^{(\infty)}\right)\mathbf{C}\right]=\mathit{tr} \left[ \mathbf{C}^T\left(\mathbf{\Pi} e^{-t\mathbf{L}^{(\tau)}} - (\mathbf{p}^{(\infty)})^T\mathbf{p}^{(\infty)}\right)\mathbf{C}\right] \, ,
\end{equation}
where $\mathbf{\Pi}$ is the diagonal matrix containing $\mathbf{p}^{(\infty)}$ on the diagonal, $\mathbf{L}^{(\tau)}$ is the above defined random walk Laplacian on the hypergraph and $(\mathbf{p}^{(\infty)})^T\mathbf{p}^{(\infty)}$ is the matrix whose $(i,j)$ entry is given by $p_i^{(\infty)}p_j^{(\infty)}$. An overview and thorough introduction to Markov stability can be found in~\cite{DSYB}.

Markov Stability $r(t;\mathbf{C})$ is a quality function quantifying the goodness of the partition $\mathbf{C}$ as a function of the time horizon of the random walk. It can be used to rank partitions of a given graph at different time scales or, alternatively, as an objective function to be maximised for every time $t$ in the space of all possible partitions of the hypergraph. We consider the latter and focus on a standard random walk on the weighted projected network defined by $K_{ij}^{(\tau)}$, allowing us to use standard optimisation algorithms developed in~\cite{DSYB,SDYB}.

\begin{remark}[Discrete time]
 Markov stability for random walks in discrete time is equivalent to the modularity of the weighted adjacency matrix $K_{ij}^{(\tau)}$, when $t=1$.
\end{remark}

\section{Applications}
\label{sec:ex}

In this section, we investigate the effect of $\tau$ on communities uncovered in artificial and real-life networks.

\subsection{Toy network}
\begin{figure}[ht]
\centering
\includegraphics[scale=0.35]{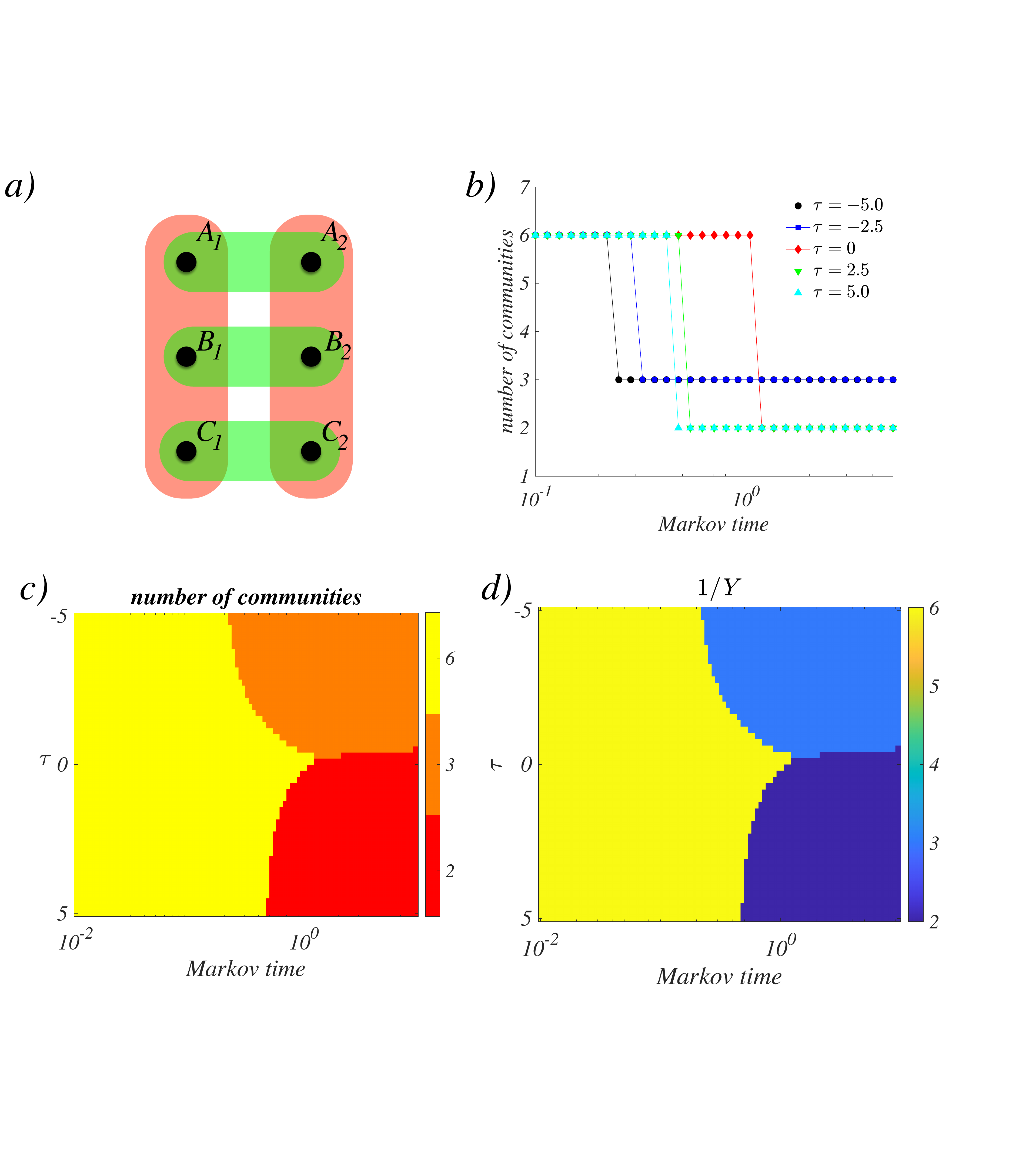}
\vspace{-1.5cm}
\caption{\textbf{Two features hypergraph}. Panel a): The hypergraph is made by $6$ nodes and $5$ hyperedges, two of size $3$ (red ones) and three of size $2$ (green ones). The former can be thought to represent the feature ``nodes with the same number'' while the former ``nodes with the same letter''. Panel b): the number of communities versus the Markov time for several values of the size bias parameter $\tau$; as expected from the theory, for large Markov time and large positive $\tau$ the SF detects $2$ communities, i.e. the two $3$-hyperedges, while for small negative $\tau$ the method reports $3$ communities, i.e. the three $2$-hyperedges. Panel c): a global view of the transition between the number of detected communities as a function of $\tau$ and the Markov time. Panel d): we report the inverse of the Deridda-Flyvbjerg number as a function of $\tau$ and the Markov time, namely a proxy of the composition of the communities.}
\label{fig:toyhyp}
\end{figure}

The first example is a toy model, a hypergraph where nodes are endowed with two features, letters $A$, $B$ or $C$, and numbers $1$ or $2$. The hypergraph is composed of $6$ nodes, $A_1$, $A_2$, $B_1$, $B_2$, $C_1$ and $C_2$, and $5$ hyperedges, $3$ of size $2$  connecting  nodes with the same letter, and 2 of size $3$ connecting nodes with the same number (see panel a) of Fig.~\ref{fig:toyhyp}). The matrix $\mathbf{K}^{(\tau)}$ is easily computed as hyperedges of size $2$ contribute a weight of $1^\tau$ in the weighted network, and hyperedges of size $3$ a weight of $2^\tau$. The transition probabilities are thus given by:
\begin{equation*}
 T_{X_1,X_2}=\frac{1}{2\times2^\tau+1}\, , T_{X_1,Y_1}=\frac{2^{\tau}}{2\times2^\tau+1}\, ,T_{X_1,Y_2}=0\, ,T_{X_1,X_1}=0\, ,T_{X_2,X_2}=0\quad \forall X,Y\in\{A,B,C\}\, ;
\end{equation*}
thus
\begin{equation*}
 \lim_{\tau\rightarrow \infty}T_{X_1,X_2}=0\, ,  \lim_{\tau\rightarrow \infty}T_{X_1,Y_1}=1/2\quad \forall X,Y\in\{A,B,C\}\, ,
\end{equation*}
namely for large $\tau$, hops among nodes with the same number are strongly favoured and hence the walker will remain for a long time in the same $3$-hyperedge. On the other hand,
\begin{equation*}
 \lim_{\tau\rightarrow -\infty}T_{X_1,X_2}=1\, ,  \lim_{\tau\rightarrow -\infty}T_{X_1,Y_1}=0\quad \forall X,Y\in\{A,B,C\}\, 
\end{equation*}
and the walker will thus spend longer periods of times in the $2$-hyperedges. In the first case, an optimisation of Markov stability is expected to return $2$ communities for sufficiently large Markov times, while  it will produce $3$ communities in the second case, as can be seen in  Fig.~\ref{fig:toyhyp}. In panel b), we report the number of communities  in the optimal partition as a function of the Markov time for several values of $\tau$. The results, combined in panel c), reveal a qualitative change of the optimal partition depending on $\tau$, as expected.

The number of communities only provides incomplete knowledge about a partition. We complement it by  the Deridda and Flyvbjerg number~\cite{DeriddaFlyvbjerg}, also known as the Simpson diversity index, defined as
\begin{equation}
\label{eq:DF}
Y=\sum_{i=1}^M\frac{S_i^2}{N^2}\, ,
\end{equation}
where $S_i$ is the number of elements in the $i$-th group.
The quantity is a standard measure of how uniform a probability is distributed and ranges from $Y=1$ when all the nodes belong to one single group to $Y=1/N$ when there are  $M=N$ groups, each one containing a single node. If the nodes are uniformly shared among the $M$ groups, i.e. $S_i \sim N/M$, then $Y\sim 1/M$.

For each couple $\tau$ and Markov time, we report (see panel d) in Fig.~\ref{fig:toyhyp}) the value of $1/Y$. Here $N=6$ and $M=2$ or $M=3$ (excluding the trivial case $M=6$); in the former case, assuming the nodes to be equally shared among the two communities, we get $Y=2 \times (3/6)^2=1/2$, while in the latter case we obtain $Y=3 \times (2/6)^2=1/3$. One can observe that indeed $1/Y = 2$ (dark blue zone) for the same set of values for which the number of communities equals $2$, and $1/Y=3$ (light blue zone) for parameters corresponding to $3$ communities.

To conclude this section, we investigate whether the size $k$ of a hyperedge increases or decreases its likelihood to be inside a community, depending on $\tau$.
The results reported in Fig.~\ref{fig:toyhyp} show a sharp transition at $\tau=0$ corresponding to a structural change in the communities detected; for negative $\tau$ the random walker will spend more time in the small size hyperedges and thus the nodes will be partitioned into communities of size $2$ corresponding to the $2$-hyperedges, i.e. $c_1=\{A_1,A_2\}$, $\{B_1,B_2\}$ and $\{C_1,C_2\}$. On the other hand if $\tau>0$ the large size hyperedges will ``capture'' the walker for long times and thus the network will be partitioned into large size communities associated to the $3$-hyperedges, i.e. $\{A_1,B_1,C_1\}$ and $\{A_2,B_2,C_2\}$.

\subsection{Hierarchical network}

\begin{figure}[ht]
\centering
\includegraphics[scale=0.17]{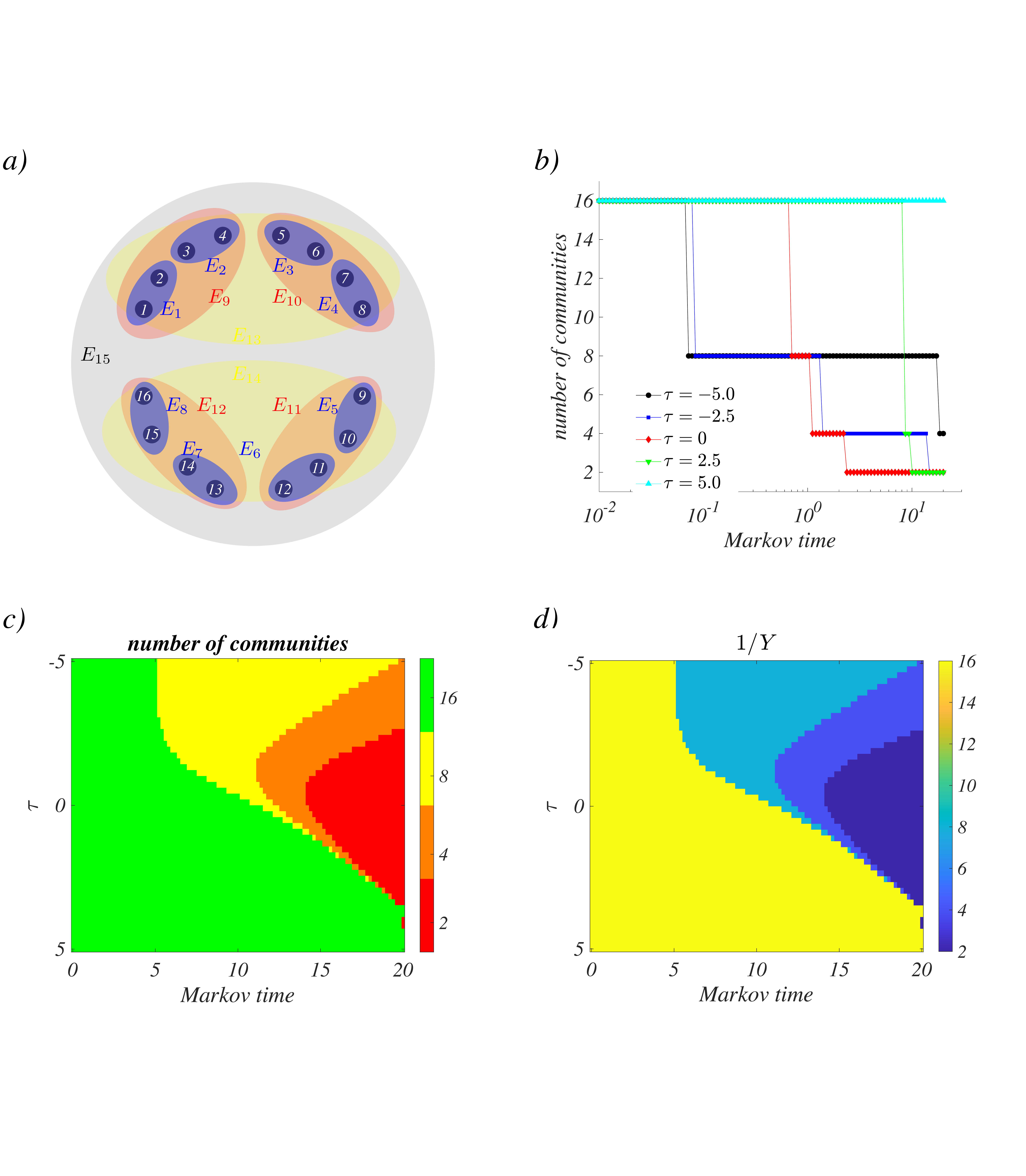}
\vspace{-1.5cm}
\caption{\textbf{Hierarchical hypergraph}. Panel a): The hypergraph is made by $16$ nodes and $15$ (non simple) hyperedges. There are $8$ hyperedges of size $2$ (blue), $4$ hyperedges of size $4$ (red), $2$ hyperedges of size $4$ (yellow) and $1$ hyperedge of size $16$ (grey). Panel b): the number of communities versus the Markov time for several values of the size bias parameter $\tau$. Panel c): a global view of the transition between the number of detected communities as a function of $\tau$ and the Markov time. One can observe that for large positive $\tau$ the SF exhibits a sharp transition between $16$ communities, each made by a single node, to $2$ communities corresponding to the two $8$-hyperedges. On the other hand negative $\tau$ allow to explore the intermediate structures passing from $2$-hyperedges, $4$-hyperedges and then $8$-hyperedges. Panel d): the proxy for the composition of the communities, $1/Y$.}
\label{fig:hierhg}
\end{figure}

Our second example  is directly inspired by the weighted hierarchical network presented in Fig. 1 of~\cite{LambiotteEtAl2014}. The hypergraph contains $16$ node and $15$ hyperedges (see panel a) Fig.~\ref{fig:hierhg}) and it has an hierarchical structure in terms of hyperedges. More precisely, $8$ hyperedges (in blue in the Figure) contain $2$ nodes, hence $4$ hyperedges (in red in the Figure) of size $4$ are made, each one obtained by merging $2$ hyperedges of size $2$. Then the process is repeated, $2$ hyperedges (in yellow in the Figure) of size $8$ are created each one containing all the nodes of $2$ hyperedges of size $4$. And finally a large hyperedge (in grey in the Figure) of size $16$ is made with all the nodes. Let us observe that the projected network is a complete network made by $16$ nodes.

We then optimise the Markov stability to determine, as a function of the Markov time, $t\in [10^{-2},10]$, and the size bias parameter $\tau$, the optimal partition of nodes into communities. The results presented in the panel b) of Fig.~\ref{fig:hierhg} clearly show that the method is able to detect the hierarchical structure of the communities, that is starting with $16$ isolated nodes for very short Markov time, the method is able to capture the intermediate coarse grained structures made by $8$, $4$ and $2$ communities as the Markov time increases.  A straightforward application of the definition~\eqref{eq:khij} allows to obtain the following values for the matrix $\mathbf{K}^{(\tau)}$:
\begin{eqnarray*}
 K^{(\tau)}_{12}&=&1+3^\tau+7^\tau+15^\tau\, , K^{(\tau)}_{13}=K^{(\tau)}_{14}=3^\tau+7^\tau+15^\tau\\
 K^{(\tau)}_{15}= K^{(\tau)}_{16}&=& K^{(\tau)}_{17}= K^{(\tau)}_{18}=7^\tau+15^\tau\text{ and }K^{(\tau)}_{1j}=15^\tau\quad \forall j=9,\dots, 16\, ,
\end{eqnarray*}
the idea being that the larger is the second index, $j$, the smaller is the number of hyperedges containing both $i$ and $j$, but with increasing sizes. For instance $i=1$ and $j=2$ belong to $4$ hyperedges whose sizes are $2$, $4$, $8$ and $16$, while $i=1$ and $j=5$ belong to $2$ hyperedges whose sizes are $8$ and $16$. By symmetry on can easily compute all the remaining entries $K^{(\tau)}_{ij}$.

Then recalling the definition of the transition probability~\eqref{eq:Tij4} one can prove that in the limit of extremely large $\tau$ we get:
\begin{equation*}
 \lim_{\tau\rightarrow \infty}T^{(\tau)}_{ij}=\frac{1}{15}\quad \forall i,j\in\{1,\dots, 16\}\, ,
\end{equation*}
that is the random walker executes jumps among nodes with uniform probability, regardless of the hyperedge size. On the other hand we can prove that
\begin{equation*}
 \lim_{\tau\rightarrow -\infty}T^{(\tau)}_{12}=1\text{ and } \lim_{\tau\rightarrow -\infty}T^{(\tau)}_{1j}=0\quad \forall j\in\{1,\dots, 16\}\setminus \{2\}\, ,
\end{equation*}
and similarly for nodes belonging to the {remaining} $2$-hyperedges. Hence the process forces the walker to remain in the smallest hyperedges, i.e. those of size $2$.

Starting from these observations, we can explain the results reported in Fig.~\ref{fig:hierhg}. In Panel c) we present the number of communities detected for each couple $\tau$ and Markov time; one can observe that for large positive $\tau$, the algorithm returns the finest partition, i.e. made of $16$ communities (green zone), up to long Markov time and then suddenly  the coarsest one made by $2$ large communities (red zone). This result is intuitive as, for large $\tau$, the transition probability are uniform. Only negative $\tau$ allow to explore the intermediate partitions (yellow and orange zones) and eventually end up with the partition into $8$ groups of size $2$, as predicted by the behaviour of the transition probability in the limit $\tau\rightarrow -\infty$. 

The information about the number of communities is complemented with the diversity index  in panel d) of Fig.~\ref{fig:hierhg}. The value $1/Y=2$ (dark blue zone) corresponds to $M=2$ communities each made by $8$ nodes, indeed $Y=2\times (8/16)^2=1/2$, the value $1/Y=4$ (light blue zone) is associated to $M=4$ communities each made by $4$ nodes, $Y=4\times (4/16)^2=1/4$. Finally the region associated to $1/Y=8$ (cyan zone) corresponds to $M=8$ and $2$ nodes per group, $Y=8\times (2/16)^2=1/8$.


\subsection{Animal hypernetwork}

\begin{figure}[ht]
\centering
\includegraphics[scale=0.15]{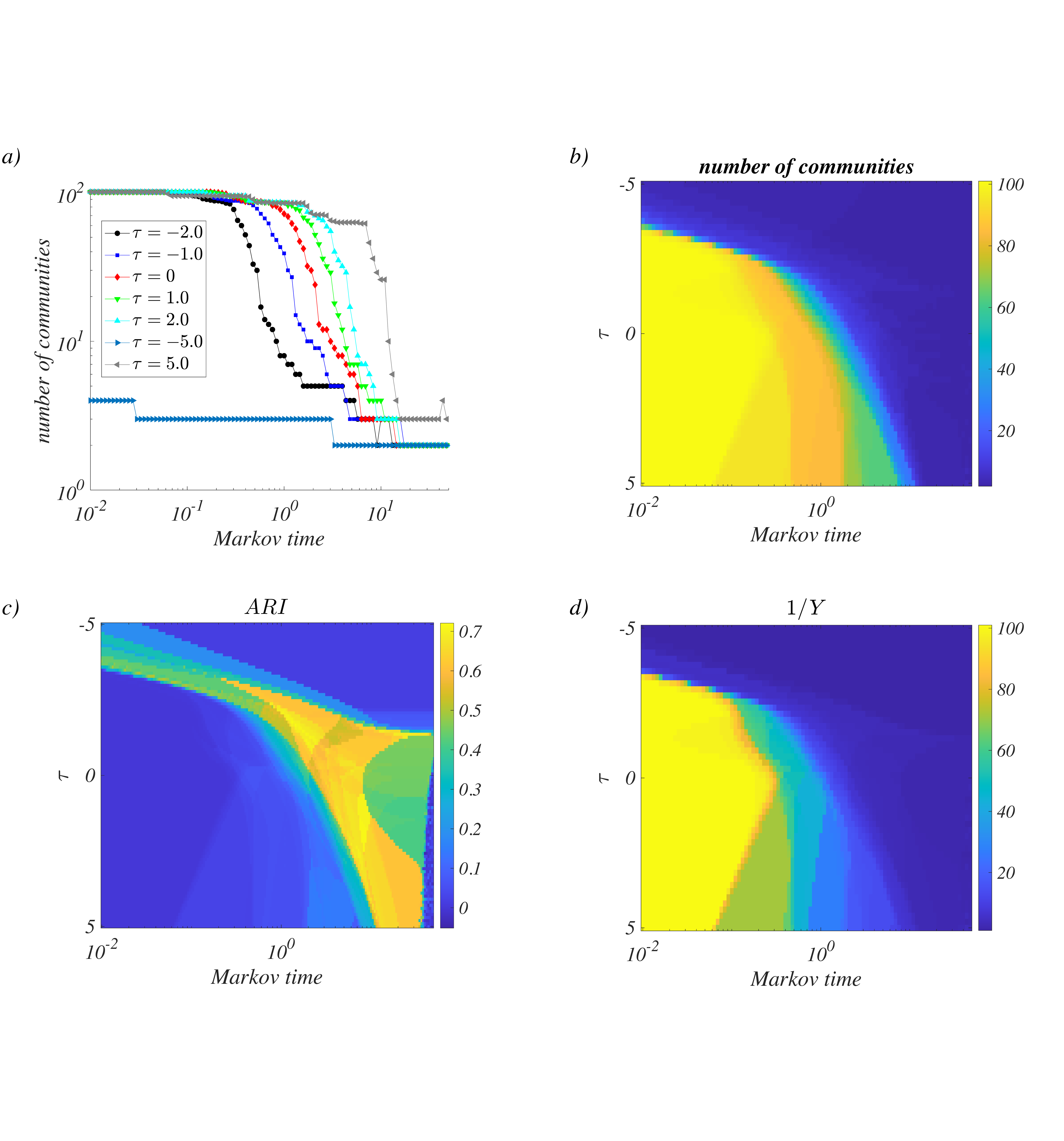}
\vspace{-1.5cm}
\caption{\textbf{Zoo database}. Panel a): number of communities versus Markov time for several values of $\tau$. Panel b): number of communities as a function of $\tau$ and Markov time. Computed communities versus the ground truth measured using the ARI, larger values are associated to a good matching among the two partitions.
Panel d): composition of the communities  as a function of $\tau$ and Markov time}
\label{fig:zoohg}
\end{figure}

The third example is based on the dataset containing an ensemble of animals from a zoologically heterogeneous set, taken from the UCI Machine Learning Depository~\cite{Dua:2019}. The dataset is made of $101$ animals, each one endowed with $20$ features, such as tail, hair, legs and so on~\cite{CarlettiEtAl2020}. For each animal we know the ground truth, that is its corresponding class, e.g. Mammal ($41$ elements), Bird ($20$ elements), Reptile ($5$ elements), Fish ($13$ elements), Amphibian ($4$ elements), Bug  ($8$ elements) and Invertebrate ($10$ elements). Here nodes are animals and hyperedges features; the goal is to use the random walk to cluster ``similar animals'' into communities. In Fig.~\ref{fig:zoohg} we report the community structure obtained by optimising Markov stability as a function of the size bias parameter $\tau$ and the Markov times. On the panel a) we report the number of communities as a function of the Markov time for several values of $\tau \in \{-5,-2,-1,0,1,2,5 \}$, while in the panel b) we present a more global view. From both panels one can appreciate the nonlinear interplay between $\tau$ and the number of communities detected at a given Markov time; indeed for small enough (negative) $\tau$, the number of detected communities is small, for all the considered Markov times. On the other hand, for a fixed Markov time, increasing $\tau$ allows to sharply pass from few to many communities, whose sizes are presented in the panel d).

Finally in panel c), we compare the community structure obtained for given $\tau$ and Markov time, with the ground truth, i.e. the known classes to which any animal belongs to. To do this we used the {\em Adjusted Rand Index} (ARI)~\cite{HA1985}, a method allowing to compare two partitions of the elements of a given set. The larger is the index the more similar are the two partitions. The ARI is an improvement of the Rand index adjusted for the chance grouping of elements. Our numerical results reveal that optimal values of ARI are obtained for complex combinations of the size bias parameter $\tau$ and Markov time, hence motivating the possibility to tune these parameters in real-world settings.

%
%

\section{Conclusions}
\label{sec:conc}

In this article, we have considered a family of random walks on hypergraphs with a parameter controlling the bias of the dynamics towards hyperedges of low or high size. As we have shown, the process naturally provides different ways to project hypergraphs on networks and includes standard approaches like the clique-reduced multigraph as special cases. The resulting projections may radically differ depending on the size bias parameter and we have explored this dependency  through its effect on community structure. Building on  Markov stability, we have developed a general framework to uncover communities in hypergraphs, which we have tested on artificial and real-world networks. For future research, interesting questions include the determination of appropriate values of the size bias parameter in empirical data, as well as a more thorough comparison of the weighted projections, for instance with graph distance measures \cite{donnat2018tracking}.

\bibliography{bib_HRW}

\end{document}